\documentclass[12pt,a4paper]{article}
\usepackage{amsmath}
\usepackage{graphicx}
\usepackage{ulem}
\usepackage{times}
\usepackage{cite}
\begin{document}
	\begin{titlepage}
		\title{Rational form of amplitude and its asymptotic  factorization}
		\author{ S.M. Troshin, N.E. Tyurin\\[1ex]
			\small   NRC ``Kurchatov Institute''-- IHEP\\
			\small   Protvino, 142281, Russian Federation}
		\normalsize
		\date{}
		\maketitle

		\begin{abstract}
		We provide  arguments for the use of the rational form of unitarization, its relation with the diffraction peak shrinkage  and asymptotics of the inelastic cross--section.
		  The  particular problems of the Regge model  and  the exponential form of unitarization with a factorized eikonal   are discussed as well.  Central role  belongs to the asymptotic amplitude factorization resulting from  Mandelstam analyticity and its  symmetry over the scattering variables.
		\end{abstract}
	\end{titlepage}

\section{Introduction.}
The {partial} elastic scattering amplitude $f_l(s)$ decreases as a linear exponent  { in} $l$ at large values of { orbital momentum} $l$ { and the Mandelstam variable} $s$  in accordance  with the  Froissart--Gribov projection formula  \cite{pdbc,frs,grb}:
\begin{equation} \label{asm}
	f_l(s)\simeq  g(s)\exp(-\mu\frac{2l}{\sqrt{s}}),
\end{equation}
where $g(s)$ is a complex function of { the} energy and $\mu$ is related to  position of the lowest $t$--channel singularity  determined by the mass of two pions.
Eq. (\ref{asm}) originates from Mandelstam representation and is consistent  with the rigorously proven domain of the amplitude analyticity  in the Lehmann--Martin ellipse \cite{mt}.  

This is why  Eq.(\ref{asm}) should be taken into account
under construction of the scattering amplitude. Actually such an amplitude differs significantly when predicting behavior of the experimentally observed integral cross--sections, slope parameter etc. Ultimately, the use of ``correct'' amplitude is important when building the hadron interaction picture. { We discussed separate aspects of the ``correct'' amplitude construction e.g. in \cite{perf,edge,crr,pl17, part21, pl21} and are attempting to provide a more extended and coherent picture of soft hadron interactions here.}

To visualize  the hadron scattering it is convenient to use a semiclassical picture and the impact parameter representation 
(cf. \cite{halz,webb, bl,gold}) since 
the impact parameter $b$ { ($b=2l/\sqrt{s}$)}  is a conserved quantity at high energies and  the scattering amplitude    is determined by the Fourier--Bessel transformation of the amplitude $F(s,t)$. 
Large--$b$ behavior of the scattering amplitude is obtained straightforwardly  from Eq. (\ref{asm}) \cite{bl,gold,schr} with substitution $b=2l/\sqrt{s}$ leading to factorization in $s$ and $b$ variables.

 In the Regge model with linear trajectory $\alpha(t)=\alpha(0)+\alpha'(0)t$ there { are} several 
unpleasant complications.
The linear form  implies Gaussian dependence on the impact parameter $b$ in contradiction with Eq. (\ref{asm}). { Such} dependence cannot be also reduced to a factorized form at large values of $b$ as required by Eq. (\ref{asm}). 

To cure the first  issue, it has been proposed to consider  nonlinear Regge trajectories (see \cite{lsl} and references therein) based on  {  accounting} for the $t$--channel two--pion loop contribution \cite{pmp,gol}. However, the problem with asymptotical factorization  {remains}. It should be noted that 
extension of Eq. (\ref{asm}) to the whole region of the impact parameter variation leads  to a factorized amplitude $F(s,t)$ which violates  unitarity in { the} $s$ and $t$ reaction channels if the total cross--section does not decrease with energy at $s\to\infty$ \cite{grb}. It is also inconsistent  with the experimentally observed shrinkage of the diffraction cone and some other experimentally measured characteristics.

We  { are discussing} a way to avoid the indicated problems.   {The reference here to the Regge model is relevant because} this model generated  a set of notions and formulas which are used for the data analysis disregarding  the respective limitations. 

 Unitarization { is the tool that}
reproduces factorization of the scattering amplitude at large impact parameter values.  This approach emphasizes the fact that unitarity plays an essential {\bf dynamical} role at all the energy values  redistributing somehow probabilities of the elastic and inelastic events under  hadron collisions. Account for Eq.(\ref{asm}) requirements (next section) provides us with the amplitude  and conclusions other than the predictions of the
 Regge model or some 
other models. This highlights the dynamical role of  unitarization and importance of the asymptotic amplitude factorization resulting from the symmetry of  Mandelstam analyticity over { the} scattering variables.
 \section{Construction of the input amplitude} 
 Unitarization itself is a well known technique to achieve consistency with the basic principle of unitarity. The question is how to choose the right form of the amplitude especially in view of the recent LHC observations \cite{dremin}.
 We  construct the input amplitude  based on its known asymptotics in the region of large $l$--values, Eq. (\ref{asm}) and use  a rational representation {for its subsequent unitarization}
  ($U$-matrix form of the scattering amplitude, see for details \cite{pl21}). We neglect  spin degrees of freedom in what follows.
  The reasons to use { the}  rational unitarization instead of { the} exponential one (eikonal) are evident. Despite simplicity of the rational form, it retains an initial type of singularity after unitarization while the exponential form transforms a simple singularity of the phase function (eikonal) into an essential singularity in the resulting amplitude \cite{bl}. It is a convinient tool for the resonance studies \cite{anis}.
  The eikonal unitarization does not allow to explain naturally the observed exceeding the black disk limit \cite{alkin1} since it corresponds to a maximal absorption at $s\to\infty$ and $b$-fixed. 
   Eikonal  leads to problems\footnote{It has been shown that experimental data imply impact parameter dependence of the logarithmic derivative of the eikonal phase \cite{alber}.} with the experimental data description.

 The scattering amplitude $f(s,b)$ relation with the input quantity $u(s,b)$   in the impact parameter representation has a simple form:
\begin{equation}\label{um}
f(s,b)=u(s,b)/[1+u(s,b)]
\end{equation}
and can be inverted 
\begin{equation}\label{umr}
u(s,b)=f(s,b)/[1-f(s,b)].
\end{equation}

At fixed  $s$ and large increasing impact parameters $b$ the factorized form with $\exp(-\mu b)$ dependence on the impact parameter $b$ should be valid for the function $u(s,b)$
since $$f(s,b)\simeq g(s)\exp(-\mu b)$$  and
\begin{equation}\label{uf}
u(s,b)/f(s,b)\to 1.
\end{equation} 
It should be noted that  factorization of $U$--matrix does not imply factorization of the scattering amplitude $f(s,b)$. 
 
We perform extension of the function $u(s,b)$ into the region of finite impact parameter values bringing violation of factorization up to  level of the amplitude $f(s,b)$  due to the rational form of  unitarization. This means mixing of the transverse and longitudinal dynamics.
Unitarization  { smoothes out the} exponential decrease of  $u(s,b)$ at small and moderate values of $b$ at finite energies, i.e. it makes $f(s,b)$ flattening in the transverse space. 

To invoke a hint of the energy dependence of $u(s,b)$, we  turn now to its energy behavior  at $s\to\infty$ and fixed $b$.
Postulating saturation  of the unitarity limit, i.e. $f\to 1$ at $s\to\infty$ at fixed $b$, we obtain that $u(s,b)\to\infty$ in this limit, Eq. (\ref{umr}). 
This assumption corresponds to the principle of maximal strength of strong interactions proposed long ago by Chew and Frautchi \cite{chew}. They noted that a `` characteristic of strong interactions is a capacity to `` saturate'' unitarity condition at high energies''.
	The behavior $u(s,b)\to\infty$ at $s\to\infty$ and $b$--fixed reflects application of this principle. Due to power-like boundedness of the $U$--matrix \cite{echa} the above  increase implies a power--like energy dependence of $u(s,b)$.

Thus, the basic principles  of the $U$--matrix construction are  Eq. (\ref{asm}) (analyticity in the cross--channel) and principle of the maximal strength of strong interactions supplemented by the assumed validity of the extrapolation of the asymptotical Eq. (\ref{asm}) into the whole region of the impact parameters variation for the input function $u(s,b)$. { The other version of the rational form of the amplitude with factor 2 in front of the input amplitude in the denominator (K-matrix) is not consistent with the principle of a maximal strength of strong interactions.}

There are also arguments for  { such a} construction based on the geometrical models of hadron collisions. 
Those  models use a  factorized form of the input amplitude: 
\begin{equation}\label{usb}
u(s,b)=g(s)\omega(b),
\end{equation}
and the dependence $g(s)\sim s^\lambda$ can  be related  { to} the effective rate of the  collision energy conversion into the mass \cite{carr,carr1,mcy}. Increase of this rate correlates with increase of { the} contribution from the newly opened inelastic channels.
The function $\omega(b)$ is interpreted as a convolution of the two matter distributions in the transverse plane, $$D_1\otimes D_2$$ of the colliding hadrons  \cite{cy}. The dependence $\omega(b)=\exp(-\mu b)$, $\mu>0$, corresponds to the energy--independent convolution of two Yukawa--type matter distributions \cite{halz} and is determined by the sizes  of  interacting particles.

 It can be concluded that  it is unitarity { that} transforms an energy--independent geometrical radius of particles into the energy--dependent  and relevant for the further consideration interaction radius $R(s)$.  The latter   determines the slope of  diffraction cone of the differential cross--section of elastic scattering and is related to the total cross--section { increase}. Thus, the unitarity plays an essential dynamical role  and is  responsible for the difference between the size of a hadron and what we call the hadron interaction region. Those two should not be equated.
\section{Energy dependence of the elastic scattering slope}
The slope of diffracion cone is defined as 
 \begin{equation}\label{bsd}
 	B(s)\equiv \frac{d}{dt}\ln \frac{d\sigma}{dt}|_{t=0},
 \end{equation}
 where the function  $B(s)$ is determined as   $$2B=\langle b^2 \rangle_{tot}\equiv {\int_0^\infty b^3db\sigma_{tot}(s,b)}/
 {\int_0^\infty bdb\sigma_{tot}(s,b)}.$$
 
 The resulting  energy dependence of $B(s)$ upon the rational unitarization is  consistent with the data and can be well described  up to the LHC energies by a power--like function \cite{epl}:
\begin{equation} \label{bs}
B(s)= B_0+As^{\lambda}
\end{equation}
 with $\lambda \simeq 0.1$.
 The transition  to the asymptotics starts  from the energy $\sqrt{s}\simeq 10^4$ GeV. This transition implies {\bf slowdown}  of the  
 $B(s)$ energy dependence towards  its asymptotic behavior which reflects presence of confinement in QCD (mass gap):  
 \begin{equation}\label{bes}
 B(s)\sim R^2(s)\sim\ln^2 s.
 \end{equation}
  The apparent speeding up of the slope shrinkage at the LHC energies can be treated as   its power--like increase in accordance with Eq. (\ref{bs}).  { The quantities $\sigma_{tot}(s,b)$ and $\sigma_{inel}(s,b)$ are related to the scattering amplitude $f(s,b)$:
  	\begin{equation} \label{sinelb}
  		\sigma_{tot}(s,b)=f(s,b), \,\,\sigma_{inel}(s,b)=f(s,b)[1-f(s,b)]
  	\end{equation}
  	and normalized as
  	\begin{equation} \label{sinelbt}
  		\sigma_{tot,inel}(s)=8\pi \int_0^\infty\sigma_{tot,inel}(s,b)bdb.
  \end{equation}}
 
 The slope $B(s)$ increase is due to spreading  the inelastic interactions into  the peripheral region of impact parameters, when
 \begin{equation}\label{bla}
 	\sigma_{tot}(s,b)\simeq \sigma_{inel}(s,b)
 \end{equation}
at large values of $b$ \cite{pl21}. 
{The Eq. (\ref{bla}) is formulated in the impact parameter space and valid in the limit of large $b$ only. For general relations see Eq. (\ref{sinelb}).}

The origin of the diffraction cone shrinkage (with transition from power--like increase of the slope to $\ln^2 s$ dependence) is related to the principle of maximal strong interactions. {This statement is also valid  for $\ln^2 s$-growth of the total cross--section and $\ln s$-growth of the inelastic cross--section at $s\to\infty$. The $\ln s$ asymptotics of the latter is due to suppression of the small impact parameter contributions because of the presence of the factor $\alpha(s,b)=1-f(s,b)$ in Eq. (\ref{sinelb}).} The principle of maximal strong interactions was formulated for the case of  real particles. In the case of virtual particles there are no rigorous arguments in favor of the above shrinkage as well as even to an existence of the similar diffraction cone  \cite{perf}.
\section{Off--shell scattering}
There is no unitarity limitation for the scattering amplitude when at least one of the initial particles is off  mass shell \cite{indur,indur1}. It does not allow one to get a bound similar to the Froissart--Martin bound. Indeed, an extension of the $U$--matrix unitarization scheme \cite{umv} for the off--shell scattering provides the following expression for the amplitude $f^*(s,b,Q^2)$ of the process 
$
h_1^*h_2\to h_1h_2:
$
\begin{equation}\label{umv}
	f^*(s,b,Q^2)=[u^*(s,b,Q^2)/u(s,b)]f(s,b).
\end{equation}
The variable $Q^2$ is  a virtuality of  the initial particle $h_1^*$. The factor $u^*/u$ can strongly change $s$-- and $b$--dependencies of the function $f$.
It is evident that there is no equation for $f^*$ similar to  Eq. (\ref{asm})   which is valid for real particles and
\[
f^*\leq u^*/u,
\]
while
$
f\leq 1
$.
There are various models in the literature for description of virtual particles scattering. One particular geometrical model has been considered in \cite{umv}. In general, one cannot make conclusion on the slope $B(s,Q^2)$\footnote{The definition for the slope $B(s,Q^2)$ is analogous to Eq. (\ref{bsd}).} in the case of virtual particles scattering since the  function $u^*$ is not known and QCD-inspired models have limited success and little justification due to the unsolved confinement problem.  

The simplest generalization of Eq. (\ref{usb}) for the case of virtual particles would be
\begin{equation}\label{usbv}
	u^*(s,b,Q^2)=g^*(s,Q^2)\omega^*(b, Q^2)
\end{equation}
with $g^*\to g$ and $\omega^*\to \omega$ at $-Q^2\to m^2$. Such factorization at large $Q^2$ can be inspired by QCD modelling of $u^*$.   However, the form of the function $\omega^*$ at large values of $b$  at $-Q^2\neq m^2$ remains to be  not known unless we resort to the models. It is due to the fact that the Mandelstam analyticity is supposed to be valid  for the real particles only. For the function $g^*(s,Q^2)$, QCD-inspired model can be useful at large values of $Q^2$, i.e. one can suppose that  $g^*(s,Q^2)\simeq 
{\tilde g}(Q^2)s^{\tilde\lambda}$ \cite{indur,indur1}. It is  just an example  demonstrating that $f^*$ at fixed $b$ and large $Q^2$ can increase like a power of energy, i.e. $f^*\sim s^{\Delta}$ at $s\to\infty$ and $\Delta\equiv {\tilde \lambda-\lambda}>0$. The latter dependence can be interpreted as a consequence of a stronger energy increase of hard diffraction (with  $\tilde \lambda$ independent of $Q^2$, see for discussions  \cite{indur,indur1}) compared to the respective growth of a soft one.
\section{Inelastic cross--section and black ring picture}
The unitarity relation can be rewritten in the form of quadratic equation for $\sigma_{tot}(s,b)$:
\begin{equation}\label{unisb}
	\sigma_{tot}(s,b)=\sigma_{tot}^2(s,b)+\sigma_{inel}(s,b)
\end{equation}
and implies that {\t dimensionless functions} $\sigma_{tot}(s,b)\leq 1$  and $\sigma_{inel}(s,b)\leq 1/4$. The latter inequality reduces the upper bound for $\sigma_{inel}(s)$ by factor of 4 compared to the upper bound for $\sigma_{tot}(s)$ 
\cite{mara}. Eq. (\ref{unisb}) allows  to generalize and improve this reduced bound 
rewriting $\sigma_{inel}(s,b)$ through  $\sigma_{tot}(s,b)$ \cite{ttwu}.

The principle of maximal strength of strong interactions  leads to the asymptotic picture of hadron interaction region in the form of  a black ring surrounding the reflective disc. This picture corresponds to the scattering matrix of elastic reaction which tends to $-1$ at fixed $b$ and $s\to\infty$ and allows one to represent the inelastic cross--section at high energies in the following form
\begin{equation} \label{sinelbe}
	\sigma_{inel}(s)\simeq 8\pi R(s) \int_0^\infty\sigma_{inel}(s,b)db=8\pi R(s)I_{inel}(s).
\end{equation}
 The dimensional integral  $I_{inel}(s)$ 
 is determined by the  { radius} of the interacting particles and
does not depend on energy \cite{edge} at $s\to \infty$. 
The asymptotic behavior of $\sigma_{inel}(s)\sim \ln s$ is evidently related to the limiting dependence of the elastic scattering amplitude in the impact parameter representation. 

Logarithmic increase of the inelastic cross-section is of interest for interpretation of the results \cite{crr} obtained for the inelastic processes   measured in the extended air showers under the cosmic rays studies \cite{horand}. The existing cosmic data suggest decreasing dependence of the ratio  $\sigma^{p-air}_{inel}(s)/ \sigma^{pp}_{tot}(s)$ up to the energy $\sqrt{s}=10^2$ $TeV$ \cite{kendi}. Despite the cosmic data are not very conclusive due to low statistics they reveal general trends with  certainty. The data do not indicate flattening of  this ratio  with the energy, i.e. the data do not exclude the possibility that the  ratio continuously decreases approaching zero at  $s\to\infty$.
 \section{Conclusion}
The objective of this note is to give arguments in favor of the rational form
of the scattering amplitude represented through the input ``amplitude''. 
We emphasize the {\bf dynamical role of unitarty} in hadron interactions and associate several experimentally observed  trends with unitarity implementation under construction of the scattering amplitude in the rational form with account of factorization and principle of the maximal strength of strong interactions used under the choice of the particular form of the input. In another words, the  choice of the input for the rational unitarization is dictated by  combination of  this principle  with the analytical properties of the scattering amplitude encoded in Eq. (\ref{asm}).
Asymptotic amplitude factorization manifests itself as a consequence of  Mandelstam analyticity  and its symmetry over scattering variables.

The above principle implies that $f(s,b)$ reaches its maximum 
value  $f(s,b)=1$ at  $s\to\infty$  and fixed $b$.
 This results in $\ln s$--dependence of $\sigma_{inel}(s)$ and $\ln^2  s$--dependence of $\sigma_{tot}(s)$ at $s\to\infty$. { Those}   are the limiting { energy} dependencies of the respective cross--sections. 
  
The experimental observation of the behavior { $\sigma_{inel}(s)\sim \ln (s)$} would be an additional proof of the principle of the maximal strength of strong interactions.
Among the  experimentally observed  trends related to unitarity we have noted 
shrinkage of the diffraction cone in elastic scattering, i.e. the respective increase of the slope $B(s)$.   
Observation of the qualitative difference in the energy dependencies of the slope $B(s)$ and those in virtual scattering $B^*(s,Q^2) $ should be treated in favor of the unitary mechanism of $B(s)$ increase. The existing data from HERA \cite{savin,lev} are scarce and not very conclusive, but imply  that the absence of any shrinkage in the case of elastic electroproduction of  J/$\Psi$ \cite{lev} can be treated in favor of considered origin of the slope  increase for  real particles.

\section*{Acknowledgements}
The correspondence on the above issues  at the early stages of the work from  Professor Evgen Martynov  is gratefully acknowledged.

\end{document}